\documentclass[prb,twocolumn,showpacs,amsmath,amssymb]{revtex4}

\usepackage{bm}
\usepackage[dvipdfmx]{graphicx}
\usepackage{color}

\begin{document}

\title{
Comparative study of Weyl semimetal, topological and Chern insulators:
thin-film point of view
}

\author{Yukinori Yoshimura$^1$}
\author{Wataru Onishi$^1$}
\author{Koji Kobayashi$^{2,3}$}
\author{Tomi Ohtsuki$^2$}
\author{Ken-Ichiro Imura$^1$}

\affiliation{$^1$Department of Quantum Matter, AdSM, Hiroshima University, Higashi-Hiroshima, 739-8530, Japan}
\affiliation{$^2$Department of Physics, Sophia University, Tokyo Chiyoda-ku 102-8554, Japan}
\affiliation{$^3$IMR, Tohoku University, Sendai 980-8577, Japan}

\date{\today}

\begin{abstract}

Regarding three-dimensional (3D) topological insulators and semimetals
as a stack of constituent 2D topological (or sometimes non-topological) systems
is a useful viewpoint.
Here, we perform a comparative study of the paradigmatic 3D topological phases:
Weyl semimetal (WSM), strong vs. weak topological insulators (STI/WTI),
and Chern insulator (CI).
By calculating the $\mathbb{Z}$- and $\mathbb{Z}_2$-indices for the thin films of
such 3D topological phases, we follow dimensional evolution of topological properties
from 2D to 3D.
It is shown that
the counterpart of STI and WTI in the time-reversal symmetry broken CI system are,
respectively, WSM and CI phases.
The number ${\cal N}_D$ of helical Dirac cones emergent on the surface of a topological insulator
is shown to be identical to the number ${\cal N}_W$ of the pairs of Weyl cones in the corresponding WSM phase: ${\cal N}_D={\cal N}_W$.
To test the robustness of this scenario against disorder,
we have studied the transport property of disordered WSM thin films,
taking into account both the bulk and surface contributions.
\end{abstract}


\maketitle

\section{Introduction}

\begin{figure*}[tbp]
\begin{tabular}{l}
(a)
\includegraphics[width=80mm, bb =0 0 360 357]{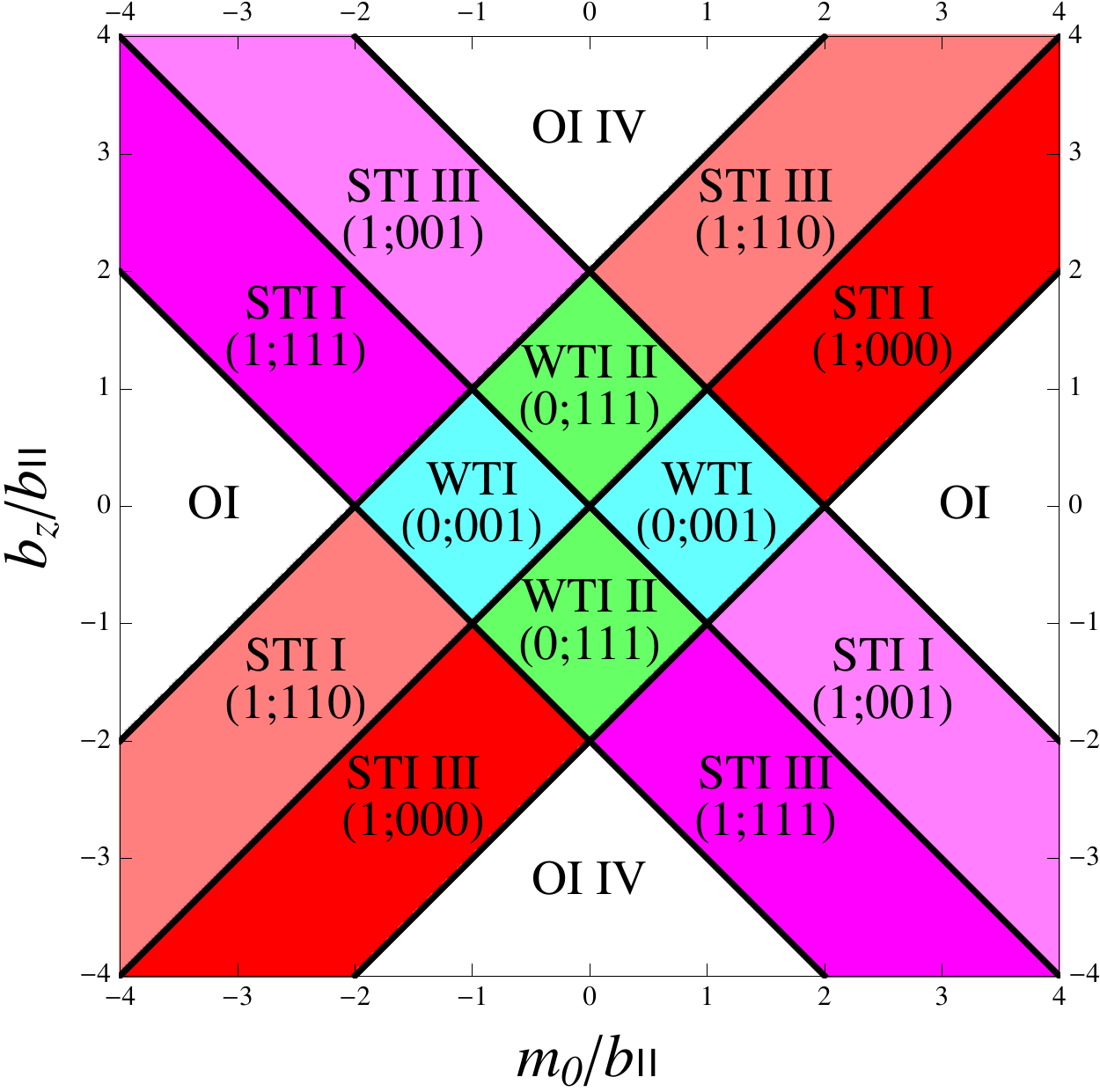}
(b)
\includegraphics[width=80mm, bb =0 0 360 357]{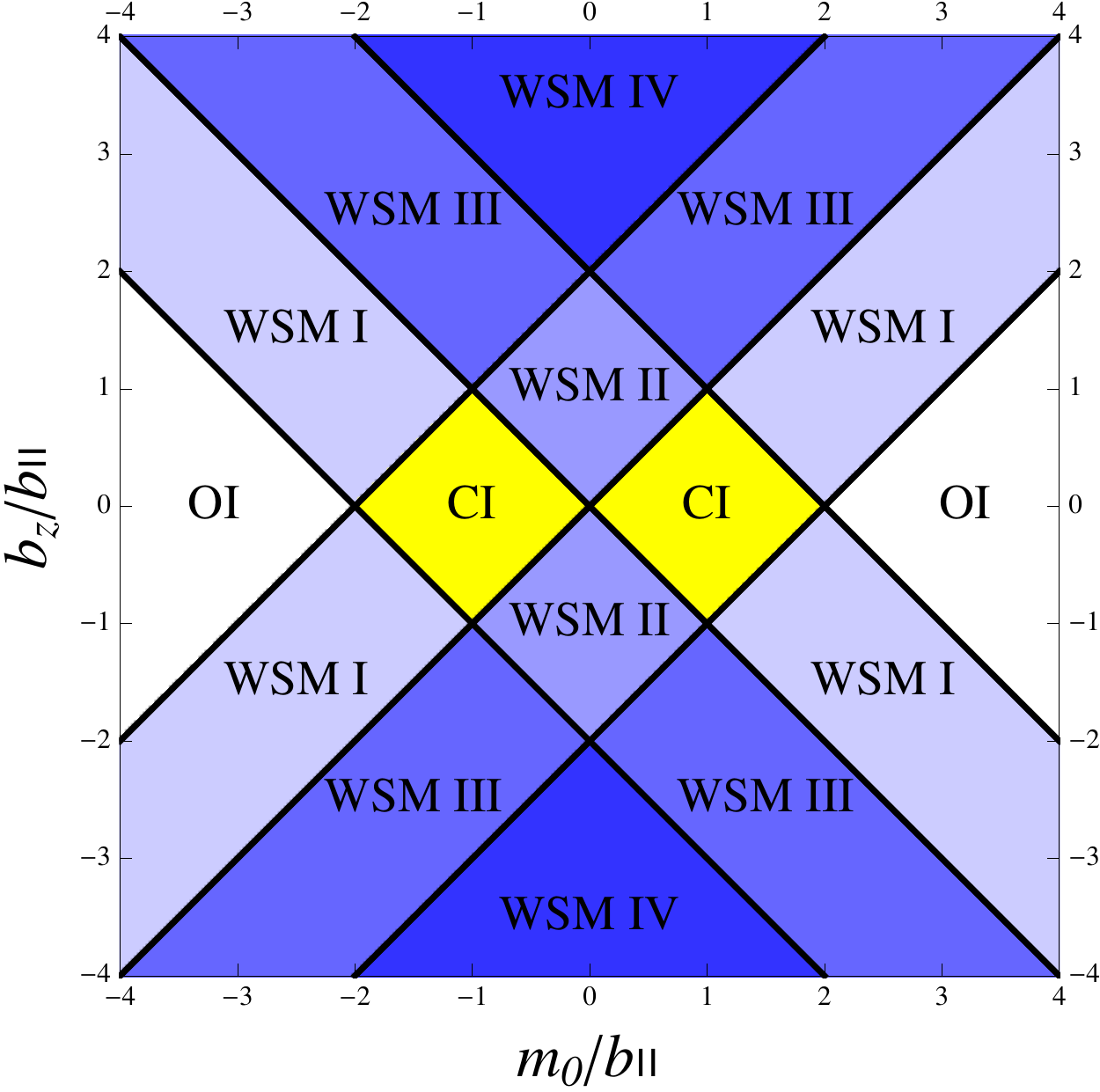}
\end{tabular}
\vspace{-2mm}
\caption{Phase diagram of 
(a) WTI/STI type model (class AII) and 
(b) CI/WSM type model (class A).
The two panels are based on a specific model defined, respectively, 
in Eqs. (\ref{H_TI}) and (\ref{H_CI}).
The Roman numbers: I, II, III, IV,
indicated in each region represent, 
in panel (a)
the number ${\cal N}_D$ of helical Dirac cones emergent on the $[001]$ surface,
while 
in panel (b)
the number ${\cal N}_W$ of the pairs of Weyl nodes. 
They are identical in WTI/STI phase and in the
corresponding CI/WSM phase.
}
\label{PD}
\end{figure*}

Weyl semimetal
is a topologically nontrivial gapless phase.
\cite{Qi_CRP} 
For theorists
it is almost trivially an interesting topic to study since it is exotic;
it has been repeatedly shown that
the existence of Weyl and Dirac cones in the band structure
results in anomalous spectral and transport properties.
\cite{Geim, Qi, Qi_CRP}
Recently, it is becoming also a target of active experimental study;
cf., its experimental discovery 
in the system of TaAs
\cite{Lv_TaAs,hasan_TaAs,Xu_TaAs} 
and NbAs.\cite{NbAs}
In this paper,
we highlight the physics of Weyl semimetal (WSM) in its relation to 
its prototypical gapped analogue,
the topological insulator 
from the viewpoint of thin-film construction.

The appearance of
Weyl points in the band structure has been 
already mentioned
in Ref. \onlinecite{SM}.
A few years later, the existence of WSM phase
has been pointed out in a pyrochlore iridate.
\cite{pyrochlore}
Recently, 
WSM has been predicted numerically 
in Ta-, Nb-, and Mo-based compounds.
\cite{bernevig, claudia,Huang_TaAs}
WSM is also shown to be robust against disorder.
\cite{KOIH, ominato, huse, syzranov, Broy, sbierski}

The 3D topological insulator is realized in a time-reversal symmetric
system. It is also under the influence of spin-orbit coupling, and belongs to symmetry class AII.
The time-reversal operator $\Theta$ of the system satisfies $\Theta^2 = -1$, 
and the topological number used for distinguishing the two types of insulating phases;
one with the surface state, the other without,
is of the $\mathbb{Z}_2$-type.
Such $\mathbb{Z}_2$-type topological insulator has also subclasses:
strong  and weak topological insulators (STI and WTI).
From a theoretical point of view
it is natural to construct such STI and WTI phases
by stacking 2D quantum spin Hall (QSH) layers.
The QSH phase can be regarded
as a 2D version of the $\mathbb{Z}_2$ topological insulator.
Such TI thin films, or stacked QSH layers
has been much studied theoretically,
\cite{shen,yoko,CXL,shen_NJP,ebi,mayu2,sacksteder15}
and experimentally as well.
\cite{TI_film,hirahara,ando1,TI_GeTe}

What happens if we stack quantum anomalous Hall (QAH) layers
instead of the QSH layers?
The resulting 3D system is typically a Chern insulator (CI), 
characterized by a $\mathbb{Z}$-type topological index, 
but it can be also a WSM; 
the coupling between the layers may turn the system to be a semimetal
[cf., typical phase diagram shown in Fig.~\ref{PD}(b)]. 
By its construction the entire class of models
breaks time-reversal symmetry and belongs to the symmetry class A.
Here, we attempt to make a close analogy between
the two types of constructions: i.e., for
(a) WTI/STI vs. (b) CI/WSM
type models.

In the two panels of Fig.~\ref{PD},
a typical phase diagram of the above two types of models
is shown.
In panel (a) WTI phases are located at the central area of topologically nontrivial region.
Topologically more robust STI phases appear 
at the periphery of such WTI phases.
Such an arrangement of the phases may seem to be natural, 
if one considers that
as varying the model parameters through WTI$\rightarrow$STI$\rightarrow$OI regions,
the number of Dirac cones emergent on the surface decreases monotonically 
as 2$\rightarrow$1$\rightarrow$0.
In this sense
STI may be regarded as partially broken WTI.
In a similar, but more definite sense specified below,
WSM can be regarded as partially broken CI.

Panel (b) of Fig.~\ref{PD} shows a typical phase diagram of the CI/WSM type model.
Here, we have adjusted the model parameters so that
it looks almost identical to panel (a), the phase diagram for the WTI/STI type model.
Indeed, they are identical except that in panel (b)
WTI and STI are replaced, respectively, with CI and WSM phases.
In CI, stacked {\it chiral} edge modes are all intact,
while in WSM,
they survive only in a part of the 2D surface Brillouin zone (BZ);
in a finite range of $k_z$ between the two Weyl points, 
they form Fermi arc surface states.
Indeed in the process of stacking
chiral edge modes
some of them become gapped in WSM.
Here, in this work
we clarify how this precisely happens
by closely examining the dimensional crossover of topological signatures
(see Fig.~\ref{zmap}).

Finally,
to test the robustness of our scenario,
we have also performed a numerical study
on the transport property of disordered WSM thin films.
Here, 
we have considered the conductance of the system
in the presence of
both the bulk and surface contributions.
The (two-terminal) conductance 
of the system 
has been calculated using the transfer matrix method,
and compared with topological index maps
(Sec. IV). 

\section{Model Hamiltonian and phase diagram}

\subsection{WTI and STI as stacked QSH layers}

The standard recipe for constructing a $\mathbb{Z}_2$-type topological insulator is 
to employ a Wilson-Dirac type effective Hamiltonian, 
\cite{CXL-1,CXL-2}
\begin{align}
 H^{\rm TI}_{\rm bulk}(\bm k)  = 
 m_{\rm 3D}(\bm k)\beta
  + \!\!\!\sum_{\mu=x,y,z}\! t_\mu\sin k_\mu \alpha_\mu, 
\label{H_TI}
\end{align}
where
\begin{equation}
m_{\rm 3D}(\bm k) = m_0 - \sum_{\mu=x,y,z} b_\mu\cos k_\mu,
\label{m_3D}
\end{equation}
is the Wilson-Dirac mass.
The anti-commuting 
$4 \times 4$ matrices $\alpha_\mu$ and $\beta$ 
can be written 
as a product of two Pauli matrices,
$\sigma_\mu$ and $\tau_\mu$,
each representing spin and orbital, 
as
\begin{equation}
\alpha_\mu = \tau_x \otimes\sigma_\mu,\:\ 
\beta = \tau_z\otimes 1_2.
\end{equation}
By changing the ratio of $m_0$ and $b_\mu$ in Eq.~(\ref{m_3D})
one can realize various different STI and WTI phases
[see phase diagram shown in Fig.~\ref{PD}(a)]. 
\cite{mayu1}
In Fig.~\ref{PD}(a) 
only the effects of uni-axial anisotropy $b_z/b_\parallel$ is taken into account, setting 
\begin{equation}
b_x=b_y=b_\parallel.
\end{equation}
Together with
strong and weak indices:
$\nu_0$ and $(\nu_1, \nu_2, \nu_3)$,
the number ${\cal N}_D$ of helical Dirac cones emergent on either the top or bottom surface 
is shown by the Roman numbers 
in each of STI and WTI phases.


The tight-binding form of the bulk effective Hamiltonian Eq.~(\ref{H_TI}) 
can be used for implementing a thin-film geometry:
\begin{align}
 &H^{\rm TI}_{\rm film}(\bm{k}_{\rm 2D})=1_{N_z} \!\otimes\!
 \left(
    m_{\rm 2D}(\bm{k}_{\rm 2D})\beta + \sum_{\mu=x,y} t_\mu\sin k_\mu\alpha_\mu
 \right)
\nonumber \\
 &-{b_z\over 2}\!
 \left(
  \begin{array}{c@{\ }c@{\ }c@{\ }c@{\,}}
   0 & 1    &      &   \\[-2mm]
   1 &\ddots&\ddots&   \\[-2mm]
     &\ddots&\ddots& 1\\
     &      & 1    & 0
  \end{array}
 \right)\!
 \!\otimes\!\beta
 +{t_z\over 2}\!
 \left(
  \begin{array}{c@{\ }c@{\ }c@{\ }c@{\,}}
   0 & -i    &      &   \\[-2mm]
   i &\ddots &\ddots&   \\[-2mm]
     &\ddots &\ddots& -i\\[0mm]
     &       & i    & 0
  \end{array}
 \right)\!
 \!\otimes\!\alpha_z,
\label{TI_film}
\end{align}
where
\begin{equation}
m_{\rm 2D}(\bm{k}_{\rm 2D}) = m_0 - \sum_{\mu=x,y} b_\mu\cos k_\mu
\label{m_2D}
\end{equation}
represents 
a 2D version of the Wilson-Dirac mass, Eq.~(\ref{m_3D}) 
for each of the constituent layers, 
while the first part of the direct products
represent the layer degrees of freedom.
In Eq.~(\ref{TI_film}) we have truncated the system in the number of stacking layers $N_z$.

Note that
in Fig.~\ref{PD}(a) 
the $b_z/b_\parallel=1$ line corresponds to 
the isotropic case, representing the 3D limit, 
while the $b_z/b_\parallel=0$ line 
represents a strong anisotropy limit, where the phase diagram becomes identical to 
that of the 2D limit.
Since $m_{\rm 2D}(\bm{k}_{\rm 2D})$ encodes topological nature
of each constituent layer,
the phase boundaries in the 2D limit are determined by the zeros of $m_{\rm 2D}(\bm{k}_{\rm 2D})$
at the four TRIM in 2D.
They are indeed at
$m_0/b_\parallel=-2, 0 , 2$.

\subsection{CI and WSM as stacked QAH layers}


As one of the 
simplest realization of WSM
with broken time reversal symmetry (class A), 
we consider the two-band
Hamiltonian of the multi-layer CI model:
\cite{ran,IT_weyl,SH,Chen15,tomi}
\begin{eqnarray}
H^{\rm CI}_{\rm bulk}({\bm k})=
m_{\rm 3D}(\bm k) \sigma_z 
+ \sum_{\mu=x,y} t_\mu\sin{k_\mu}  \sigma_\mu,
\label{H_CI}
\end{eqnarray}
where
$m_{\rm 3D}(\bm k)$ is the one given in Eq.~(\ref{m_3D}) 
and $\sigma_\mu$ are Pauli matrices. 

In parallel with the conversion from
Eq.~(\ref{H_TI}) to Eq.~(\ref{TI_film}),
the thin-film version of Eq.~(\ref{H_CI}) reads,
\begin{align}
 H^{\rm CI}_{\rm film}(\bm{k}_{\rm 2D})
 = & \ 
 1_{N_{z}} \otimes
 \left(
  m_{\rm 2D}(\bm{k}_{\rm 2D}) \sigma_z
  +\sum_{\mu =x, y} t_\mu\sin{k_\mu} \sigma_\mu
 \right)
\nonumber\\
 & -\frac{b_z}{2}
 \begin{pmatrix}
    0 &  1     &        &        \\[-1mm]
    1 & \ddots & \ddots &        \\[-1mm]
      & \ddots & \ddots &     1  \\
      &        &  1     &     0
 \end{pmatrix}
 \otimes \sigma_z.
\label{CI_film}
\end{align}
Here, the first line of the right-hand side of the equation
represents a contribution from each QAH layer.\cite{QiWuZhang} 
QAH is a 2D version of $\mathbb{Z}$ type topological insulator 
characterized by $\mathbb{Z}$ type topological index.
The Hamiltonian Eq.~(\ref{CI_film}) represents stacked QAH layers.

In the limit of $b_z\rightarrow 0$, each of such QAH layers 
decouples, representing the 2D limit of the model. 
The abscissa of the phase diagram
(the $b_z/b_\parallel=0$ line) shown in Fig.~\ref{PD}(b)
falls on this case,
exhibiting two different QAH regions:
one with $\sigma_{xy}=+e^2/h$ ($m_0/b_\parallel \in [0,2]$),
the other with $\sigma_{xy}=-e^2/h$ ($m_0/b_\parallel \in [-2,0]$).
As a finite layer coupling $b_z$ is introduced,
these two QAH phases evolve into CI phases.
These are 3D topological phases that have fully inherited the 2D-type
topological character of the constituent QAH phase;
for each
$k_z$ in the BZ, the system
shows a QAH effect,
exhibiting a gapless chiral edge mode.

Yet, as $b_z$ is increased, 
(i) a pair of Weyl points
[or (ii) two pairs of Weyl points] appear, 
depending on the value of $m_0/b_\parallel$:
e.g., for $b_z/b_\parallel=1$,
case (i) $m_0/b_\parallel\in [1,3]$ or $[-3,-1]$, while
case (ii) $m_0/b_\parallel\in [-1,1]$
in the bulk 3D BZ,
driving the system into a semi-metallic (gapless) phase.
The Weyl points appear at $k_z = \pm k_0$, with
(i) $\bar{\Gamma}$ or $\bar{M}$: $(k_x,k_y)=(0,0)$ or $(\pi,\pi)$, 
(ii) $\bar{X}$ and $\bar{Y}$: $(k_x,k_y)=(\pi,0)$ and $(0,\pi)$. 
{\it
These WSM phases can be regarded as
partially broken CI} in the sense that
in these phases
the 2D topological character of the constituent QAH layers
is only partially maintained.
In CI phases
any slice of the system at a given $k_z$ in the BZ,
say, 
the entire BZ is topologically nontrivial, 
while in the WSM phase only a part of the BZ,
$k_z\in [-k_0,k_0]$, is topologically nontrivial; 
a gapless surface, connecting the Weyl cones to form the so-called Fermi arc, appears. 


\begin{figure*}[tbp]
\begin{tabular}{l}
(a)
\includegraphics[width=150mm, bb =0 0 700 202]{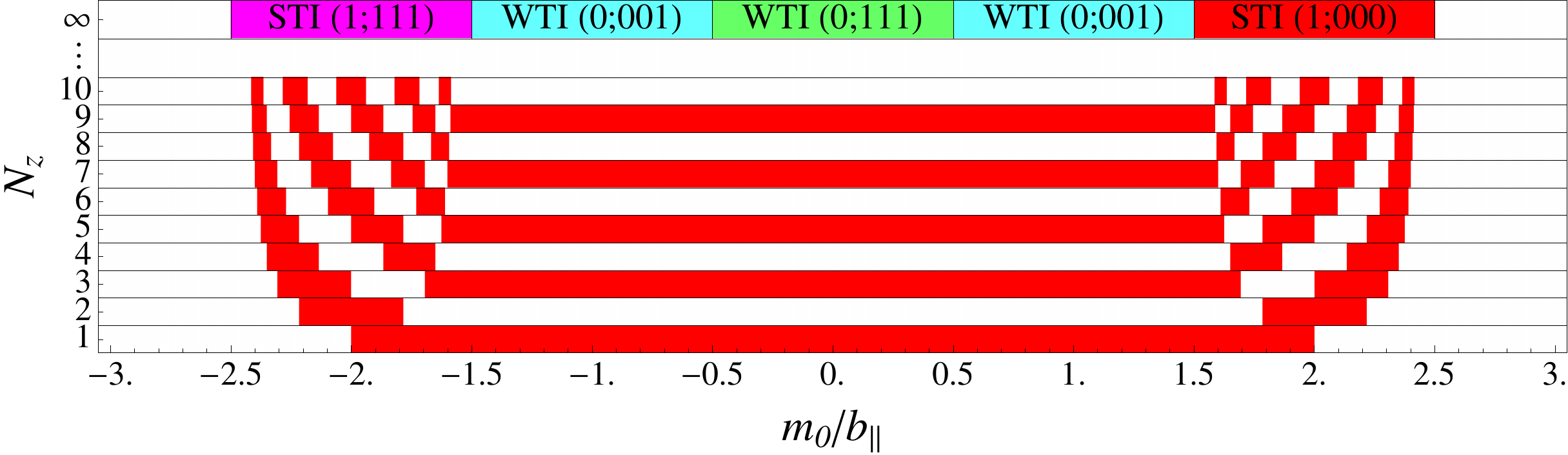}
\\
(b)
\includegraphics[width=150mm, bb =0 0 700 202]{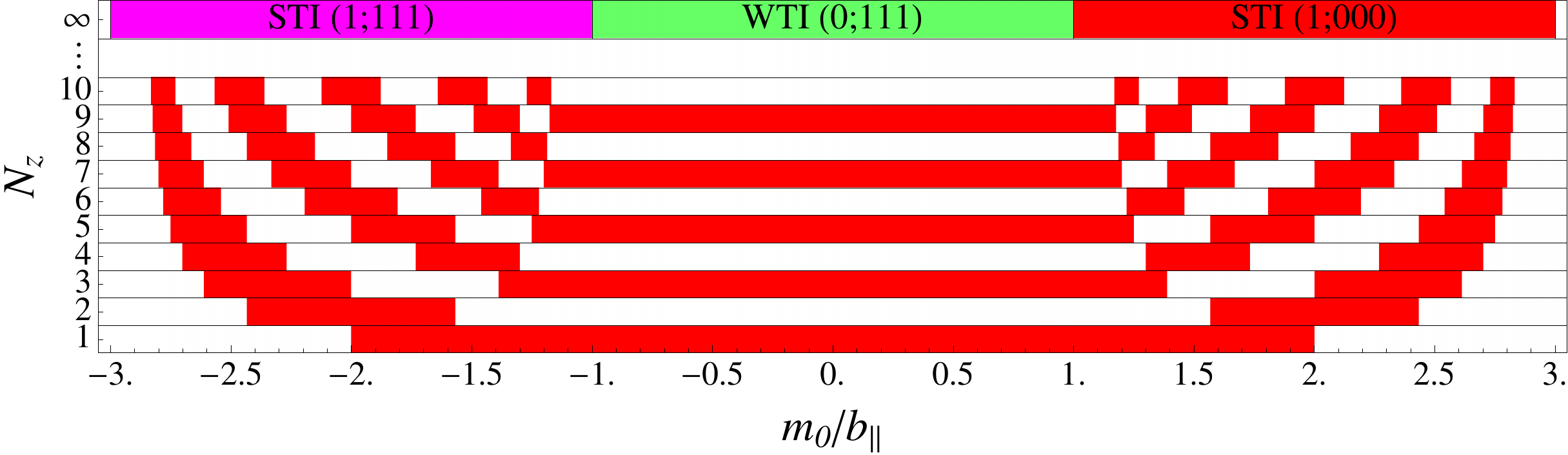}
\end{tabular}
\vspace{-2mm}
\caption{$\mathbb{Z}_2$-index map
in case of (a) an anisotropic: $b_z/b_\parallel = 0.5$ vs.
(b) the isotropic: $b_z/b_\parallel = 1$
choice of parameters.
The red [white] region corresponds to the range of QSH ($\nu=1$) [OI ($\nu=0$)]. 
The stripe pattern in the central region represents 
an even-odd feature of the 2D $\mathbb{Z}_2$ character QSH/OI, 
while a mosaic-like feature is also characteristic
at both ends of the stripe. 
The mosaic pattern corresponds to the the STI phase in the 3D limit. 
The precise nature of the mosaic pattern depends on the SOC hopping parameter $t_z$, 
here set to be $t_z/b_{\parallel} =0.5$.
}
\label{z2map}
\end{figure*}

\begin{figure*}[tbp]
\begin{tabular}{l}
(a)
\includegraphics[width=150mm, bb =0 0 545 150]{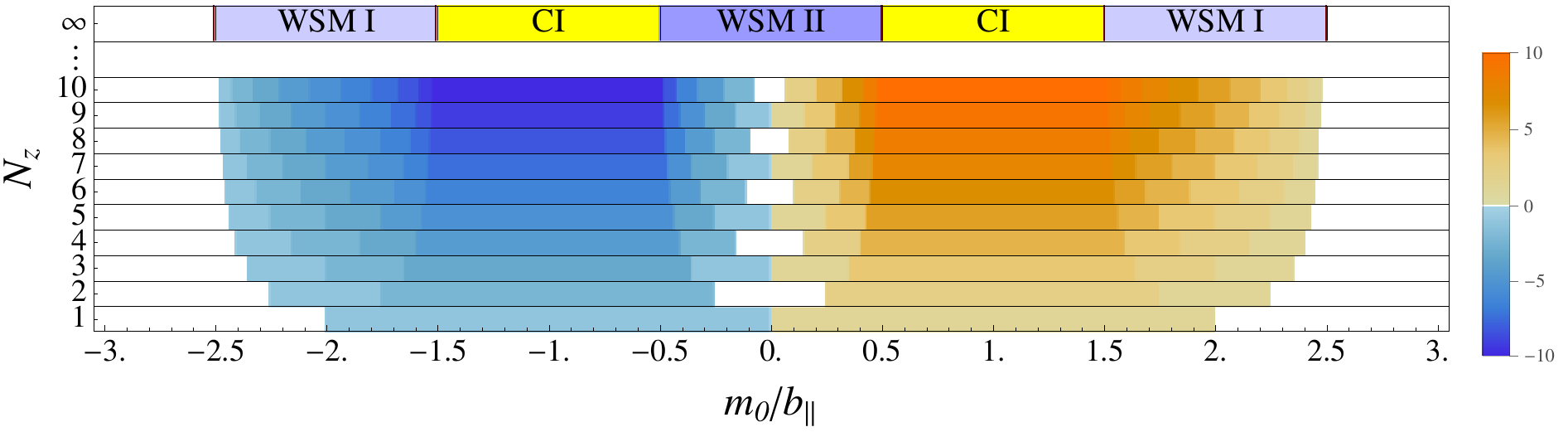}
\\
(b)
\includegraphics[width=150mm, bb =0 0 545 150]{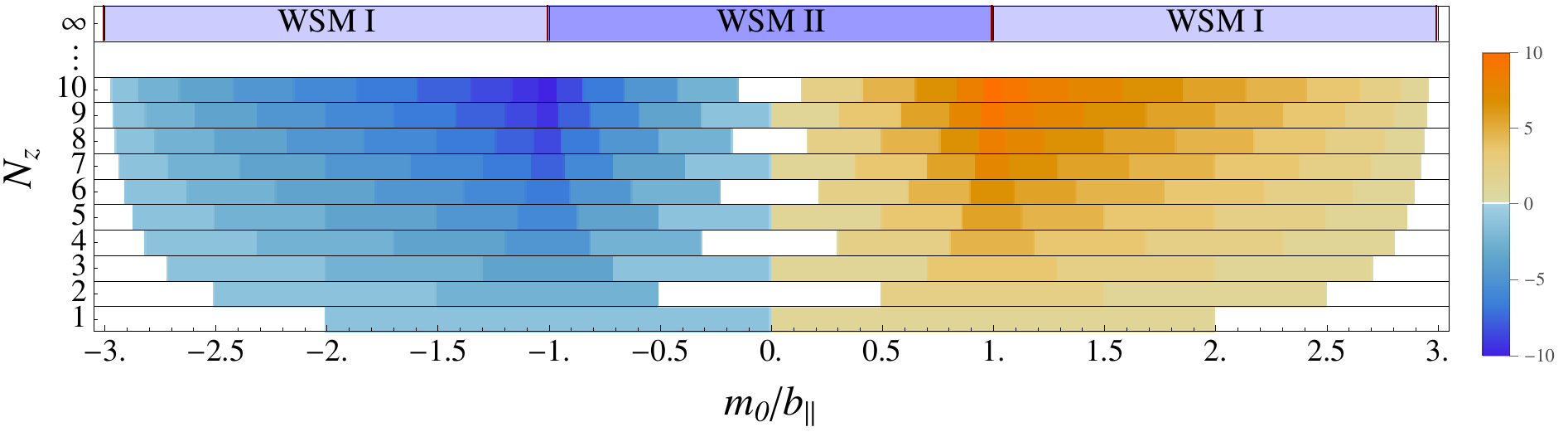}
\\
(c)
\includegraphics[width=150mm, bb =0 0 542 150]{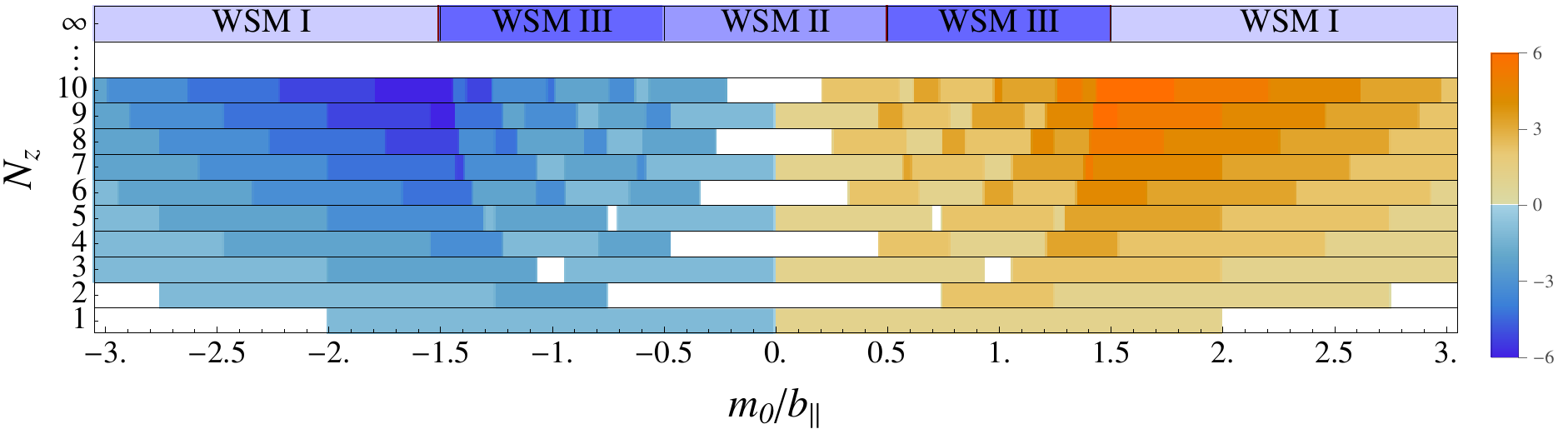}
\end{tabular}
\vspace{-2mm}
\caption{$\mathbb{Z}$-index map, representing the dimensional crossover of topological signatures
in the CI/WSM class of models (symmetry class A). 
The three panels correspond to the cases of
(a) $b_z/b_\parallel=0.5$, (b) $b_z/b_\parallel=1$, (c) $b_z/b_\parallel=1.5$.
}
\label{zmap}
\end{figure*}

As in the case of Fig.~\ref{PD}(a), 
in Fig.~\ref{PD}(b),  
the fully anisotropic line ($b_z/b_\parallel=0$) corresponds to the 2D limit of the phase diagram,
while the isotropic line ($b_z/b_\parallel=1$) corresponds to the {\it purely 3D limit}.
A thin-film situation may fall on halfway between these two limits.
Indeed,
evolution of the phases in Figs.~\ref{PD}(a) and \ref{PD}(b)
from $b_z/b_\parallel=0$ to $b_z/b_\parallel=1$
can be interpreted
as representing a dimensional crossover of the system
from 2D to 3D.
\cite{Scheurer14,cmap}
In the following section
we examine this point
more closely by explicitly studying the evolution of 2D topological signatures 
in a concrete thin-film situation
at a varying number of stacking layers.

\section{Mapping the topological index: $\mathbb{Z}_2$- and $\mathbb{Z}$-index maps}
Thin films of WTI/STI and CI/WSM can be regarded as 
2D $\mathbb{Z}_2$- and $\mathbb{Z}$- type topological insulators,
or an effective 2D QSH or QAH system.
To characterize them
we introduce and calculate
$\mathbb{Z}_2$- and $\mathbb{Z}$- topological indices \cite{BHZ, QiWuZhang, TKNN} 
adapted to such thin-film systems.

\subsection{$\mathbb{Z}_2$-index map for the WTI/STI class}

In case of the
WTI/STI type models as defined in Eq.~(\ref{TI_film}),
we can define and calculate 
a 2D $\mathbb{Z}_2$-type topological index $\nu$
as given in Ref.~\onlinecite{cmap}. 
Here, we recalculate the same index
using the convention of Eq.~(\ref{TI_film})
at different mass parameters and at different numbers of stacked layers.
Also, following the same reference,
we plot the data in the form of ``$\mathbb{Z}_2$-index map'',
in the space of 
$m_0/b_\parallel$ vs. $N_z$: the number of stacked layers.
Examples of such
$\mathbb{Z}_2$-index maps are shown in
Fig.~\ref{z2map}.

The last row ($N_z=1$) of the index map 
represents
the appearance of QSH vs.~OI phases in the 2D limit,
while the remaining part of the map shows
how this distribution evolves as a function of $N_z$.
Globally, one can recognize two types of patterns
in this map:
stripe vs.~mosaic-like irregular patterns.
The locations of the two patterns on the map
indicate roughly the positions of WTI and STI phases
in the 3D limit;
the appearance of a mosaic pattern is a precursor of STI phase in the 3D limit,
while
a stripe pattern implies the existence of WTI phase in the same limit
(see Fig.~\ref{z2map}).
While
these two patterns reflect the 
topological features in the 3D limit,
they have a slight difference in their physical origins.

The mosaic pattern
in the $\mathbb{Z}_2$-index map
is due to hybridization of the top and bottom surface wave functions
in the thin film geometry;
it represents
how the surface state wave functions penetrate into the bulk.
Though surface states of an STI and of a WTI
are localized in the vicinity of the surface,
they do have a finite penetration length.
The penetration of the surface state wave function is not necessary a simple exponential decay,
but it can show a damped oscillation.
The appearance of a mosaic pattern
in the $\mathbb{Z}_2$-index map
happens precisely in the latter case.
Naturally, the depth of penetration and the oscillation pattern
depends on the choice of parameters,
so does the resulting mosaic pattern.
In the present model
whether the surface state wave function
show damped oscillation or overdamping depends on 
the ratio $t_z/b_z$ 
(see Appendix A for details).
This surface/bulk mechanism for the mosaic pattern stands,
naturally, when the given set of parameters corresponds to a situation
in which a surface state appears on the $[001]$ surface in the 3D limit.

The physical origin of the stripe pattern is, on the contrary,
much related to the edge modes in the thin-film construction.
As far as the low-energy transport properties are concerned,
stacked QSH layers can be regarded as coupled 1D helical modes.
On one hand,
such coupled 1D modes hybridize and become gapped
when $N_z$ is even, while
a single gapless mode remains when $N_z$ is odd. 
\cite{stern,MatsuAri,takane}
This is typically the situation that underlies
a stripe pattern in the $\mathbb{Z}_2$-index map.
On the other hand,
stacked QSH layers lead to a WTI in the 3D limit,
in which
coupled 1D helical modes evolve into two helical Dirac cones
that appear on side surfaces of a WTI.
Therefore, a stripe pattern in the $\mathbb{Z}_2$-index map implies
the existence of a WTI phase in the 3D limit.

However,
precisely speaking,
this {\it edge} picture for the stripe pattern is 
justified
only when $[001]$ surface is gapped 
[this is indeed the case in the WTI $(0;001)$ phase]
in our construction.
Note that
in Fig.~\ref{z2map}
the stripe pattern in panel (a)
corresponds to the WTI $(0;001)$ phase
in the range of parameters $-1.5<m_0/b_\parallel<-0.5$ and $0.5<m_0/b_\parallel<1.5$,
while
in panel (b)
the same pattern corresponds to the WTI $(0;111)$ phase 
in the 3D limit.
In the second regime
the above criterion is no longer legitimate, but 
it still shows a stripe pattern.
In the Appendix
we show that the stripe pattern in this case
can be understood in the surface/bulk picture.
Indeed,
a complete interpretation of the two patterns in the $\mathbb{Z}_2$-index map
involves
a more thorough examination of the two scenarios:
bulk/surface vs. edge points of view
as described in Appendix A.
It is shown that the two pictures are
{\it complementary} in the formation of the two patterns.
Also, this complementarity
can be regarded as a manifestation of the guiding principle,
the bulk-boundary correspondence,
\cite{HG}
which is generic to all types of topological quantum phenomena,
here
in the formation of mosaic vs.~stripe patterns.

\subsection{$\mathbb{Z}$-index map for the CI/WSM class}

In case of the
CI/WSM type model of thin-film as defined in Eq.~(\ref{CI_film}),
we can define and characterize its thin-film
by a 2D $\mathbb{Z}$-index,\cite{TKNN}
again
at different mass parameters and at different numbers of stacked layers.
We then plot the index in the same way 
as we did for the $\mathbb{Z}_2$-type models
to form of a $\mathbb{Z}$-index map.
Some examples of the $\mathbb{Z}$-index map are shown in Fig.~\ref{zmap}.
Thanks to the notation used in
Eqs.~(\ref{H_CI}) and (\ref{H_TI}),
the similarity of the two models is explicit.
Recall also the similarity and correspondence at the level of the phase diagram
(see Fig.~\ref{PD}).

The $\mathbb{Z}$-index,
or the Chern number
is defined in terms of Berry curvature
integrated over the entire BZ.
\cite{TKNN}
Yet, a non-vanishing contribution arises only from
singularities of the band structure, 
i.e., from Dirac cones.
The so-called ``Dirac cone argument''
allows for expressing the Chern number as a sum of contributions from
such Dirac cones.
\cite{MO,HKW}
In the case of Wilson-Dirac type models on a cubic/square lattice 
the Dirac cones appear
only on the four inversion symmetric points in the BZ,
and correspondingly,
the expression for the related $\mathbb{Z}$- and $\mathbb{Z}_2$-indices
simplifies significantly.
\cite{KGC}

\subsubsection{$\mathbb{Z}$-index map, the edge point of view, comparison of the two maps}

A single layer of the CI/WSM film prescribed by Eq.~(\ref{CI_film})
is nothing but the QAH insulator studied in Ref.~\onlinecite{YIFH}.
Here, 
following Ref.~\onlinecite{YIFH}
we express the Chern number ${\cal N}$
in terms of the band index $\delta_{\bm k}=\pm 1$
at the inversion symmetric points:
$\Gamma (0,0)$, X$(\pi,0)$, Y$(0,\pi)$ and M$(\pi,\pi)$
in the following way:
\begin{equation}
{\cal N}={1\over 2}\delta_\Gamma
+{1\over 2}\delta_{\rm X}
+{1\over 2}\delta_{\rm Y}
+{1\over 2}\delta_{\rm M}.
\end{equation}
At the neighborhood of the four Dirac points ${\bm k}_{D}$ ($D=\Gamma$, X, Y, M), ${\bm k}={\bm k}_{D}+{\bm p}$,
Eq.~(\ref{CI_film}) can be expressed as follows,
\begin{align}
 H({\bm k}_{D}\!+\!{\bm p}) 
 =\tilde{t}_x p_x \sigma_x + \tilde{t}_y p_y \sigma_y 
  + \left( \tilde{m} + O(|\bm p|^2) \right) \sigma_z,
\end{align}
by using ${\bm k} \cdot {\bm p}$-approximation.
The specific values of $\tilde{t}_x$, $\tilde{t}_y$ and $\tilde{m}$ for each 
symmetric points ${\bm k}_{D}$ are shown in TABLE \ref{txtym}.
Then one can define $\delta_{\bm k}$ such that
\begin{equation}
 \delta_{\bm k} = -{\rm sgn}[\tilde{t}_x]{\rm sgn}[\tilde{t}_y]{\rm sgn}[\tilde{m}].
\end{equation}
\begin{table}[htbp]
\caption{List of $\tilde{t}_x$, $\tilde{t}_y$ and $\tilde{m}$ 
at four Dirac points ${\bm k}_{D}$}
\begin{tabular}{ccccccccccccc}
\hline\hline
${\bm k}_{D}$&&
$\tilde{t}_x$ &&
$\tilde{t}_y$ &&
$\tilde{m}$&\\
 \hline
$(0, 0)$&&
$t_x$ &&
$t_y$ &&
$m_0 - b_x - b_y$
\\
$(\pi, 0)$&&
$-t_x$ &&
$t_y$ &&
$m_0 + b_x - b_y$
\\
$(0, \pi)$&&
$t_x$ &&
$-t_y$ &&
$m_0 - b_x + b_y$
\\
$(\pi, \pi)$&&
$-t_x$ &&
$-t_y$ &&
$m_0 + b_x + b_y$
\\
\hline\hline
\end{tabular}
\label{txtym}
\end{table}
The single-layer QAH model is also a specific spin sector of the BHZ model.
Naturally, the phase diagrams of the two models resemble each other. 
In the QAH model, simply
the two QSH phases of the BHZ
are replaced with two different QAH phases with
 $\sigma_{xy}=\pm e^2/h$.
In the $\mathbb{Z}$-index map shown in Fig.~\ref{zmap}
the bottom line represents the single-layer case,
exhibiting 
two types of QAH phases with
\begin{equation}
\sigma_{xy}={\cal N}{e^2\over h}=\pm {e^2\over h}
\end{equation}
in the range of parameters, respectively,
$m_0/b_\parallel \in [0,2]$ and
$m_0/b_\parallel \in [-2,0]$.

As more layers are stacked,
contributions of different layers
accumulate,
but they also sometime interfere,
resulting in a partial or full cancellation of a non-trivial contribution.
In the presence of such interference,
the Chern number ${\cal N}$ is at most equal to $N_z$
(in magnitude).
In the $\mathbb{Z}$-index map shown in Fig.~\ref{zmap}, 
regions of larger Chern number ${\cal N}$ are indicated by a thicker color.
In the case of $\mathbb{Z}_2$-type models,
the interference
of contributions of the neighboring layers
is determined by the competition of two types of hopping terms:
$t_z$ and $b_z$ in the stacking direction.\cite{cmap}
 Here, in the $\mathbb{Z}$-type case
the $t_z$-type hopping is absent by construction, 
so that formulations developed in the Appendix of Ref.~\onlinecite{cmap}
become simpler.
The thin-film Hamiltonian $H_{\rm film}$ for the
CI/WSM case
can be made block diagonal by a simpler procedure, i.e.,
\begin{align}
 &\left(P\otimes 1_2 \right)^\dagger 
  H_{\rm film}(\bm{k}_{\rm 2D}) 
  \left(P\otimes 1_2 \right)= 
\nonumber
\\
 &\begin{pmatrix}
              H_{q_1}(\bm{k}_{\rm 2D}) &  &  & \\
    &      \! H_{q_2}(\bm{k}_{\rm 2D}) &  &    \\
    &  &   \! \ddots                   &       \\
    &  &  &\! H_{q_{N_z}}(\bm{k}_{\rm 2D})
 \end{pmatrix},
\label{CI_diag}
\end{align}
where the orthogonal matrix $P$ takes the following form;
its $(n,l)$-element is given by
\begin{equation}
P_{nl}
=
\sqrt{\frac{2}{N_z+1}}\sin{
   \left(
        \frac{n l}{N_z+1}\pi
   \right).
}
\end{equation}
Thus, contributions from different layers $l=1,2,\cdots,N_z$,
are rearranged into those from
different sectors $q_j=-(N_z+1)/2+j$ 
$(j=1,2, \cdots, N_z)$, 
and in this $q_j$-representation they decouple:
\begin{equation}
{\cal N}=
\sum_{l=1}^{N_z}
{\cal N}_{q_j},
\end{equation}
where ${\cal N}_{q_j}$ is a contribution from the ${q_j}$-th sector, determined by
the ${q_j}$-th diagonal block of the diagonalized Hamiltonian,
Eq.~(\ref{CI_diag}),
which takes the following $2 \times 2$ matrix form:
\begin{eqnarray}
H_{q_j} (\bm{k}_{\rm 2D})=
m_{q_j} (\bm{k}_{\rm 2D}) \sigma_z
+\sum_{\mu=x,y} t_\mu \sin{k_\mu}  \sigma_\mu,
\end{eqnarray}
where
\begin{equation}
m_{q_j} (\bm{k}_{\rm 2D}) = m_0 - b_\parallel \sum_{\mu=x,y}\cos{k_\mu} - b_z \cos{\frac{j \pi}{N_z + 1}}.
\label{ml}
\end{equation}
Due to the last term of Eq.~(\ref{ml}), which we will call
\begin{equation}
\Delta m_{q_j}=b_z \cos \frac{j \pi}{N_z + 1},
\label{dmn}
\end{equation}
the location of two QAH regions are shifted by the amount ${\Delta m_{q_j}\over b_\parallel}$;
i.e.,
\begin{align}
 {\cal N}_{q_j}=+1
 \quad {\rm if}\quad
 {m_0\over b_\parallel} 
 \in \left[{\Delta m_{q_j}\over b_\parallel}, 2+{\Delta m_{q_j}\over b_\parallel}\right],
\end{align}
while
\begin{align}
 {\cal N}_{q_j}=-1
 \quad {\rm if}\quad
 {m_0\over b_\parallel} 
 \in \left[-2+{\Delta m_{q_j}\over b_\parallel}, {\Delta m_{q_j}\over b_\parallel}\right]. 
\end{align}
Note that $\Delta m_{q_j}$ is the counterpart of $\lambda_{q_j}$ for the $\mathbb{Z}_2$-type model
given in Eq.~(31) of Ref.~\onlinecite{cmap}.
Thus, contributions from different sectors 
$q_j=-(N_z-1)/2, -(N_z-1)/2+1, 
\cdots,
(N_z-1)/2$,
are superposed and interfere,
resulting in a pattern of the $\mathbb{Z}$-index map
as shown in Fig.~\ref{zmap}.
The overall structure of the $\mathbb{Z}$-index map is
much similar to that of the $\mathbb{Z}_2$-index map,
except that the latter is in bicolor
while the former is in multicolor. 
Yet, apart from that
distribution pattern of different topological numbers
is much alike.
One can recognize the same two characteristic patterns:
stripe vs.~mosaic,
which was important in the interpretation of the $\mathbb{Z}_2$-index map,
here also in the $\mathbb{Z}$-index map.

In the three panels of Fig.~\ref{zmap},
an even/odd feature, or a stripe pattern appears in the central region;
here, we are referring to an even/odd feature embedded in the (globally) mosaic pattern,
somewhat analogous to the one seen in the WTI $(0;111)$ phase in Fig.~\ref{z2map_w}. 
This even/odd feature is a result of the cancellation between
${\cal N}_{q_j}$ and ${\cal N}_{-q_j}$ ($=-{\cal N}_{q_j}$),
i.e., cancellation between the edge modes of opposite chiralities
in this range of parameters.
If $N_z$ is even, this cancellation is complete, while if $N_z$ is odd,
the central term ${\cal N}_0$ remains.
The width of the ${\cal N}=0$ region becomes narrower as $N_z$ is increased
even if $N_z$ is even.
The width is determined by the last cancelling pairs,
${\cal N}_{1/2}$ and ${\cal N}_{-1/2}$,
and therefore given as
\begin{align}
   \frac{\Delta m_{-1/2}}{b_{\parallel}} 
 - \frac{\Delta m_{1/2}}{b_{\parallel}} 
 = {b_z\over b_\parallel} \sin{\pi\over 2(N_z+1)}.
\end{align}
Comparing 
$\Delta m_{q_j}$ given in Eq.~(\ref{dmn})
with Eqs.~(31) and (32) of Ref.~\onlinecite{cmap},
one can verify that
{\it
all the phase boundaries of the $\mathbb{Z}_2$-index map}
as shown in
Fig.~\ref{z2map} 
{\it reduce to those of the corresponding $\mathbb{Z}$-index map
in the limit of $t_z \rightarrow 0$.}
To make this analogy more explicit
we have added weak-like indices \cite{YIFH}
on the $\mathbb{Z}_2$-index map 
[see Figs.~\ref{z2map_w}(a) and \ref{z2map_w}(b) in Appendix].


\subsubsection{Back to the bulk-boundary relations, complementarity of the two pictures}

We have so far made a close comparison of
$\mathbb{Z}$- and $\mathbb{Z}_2$-index maps
mainly from the viewpoint of edge mechanism (at least from the $\mathbb{Z}$-side).
The edge picture we have been based on  
in the description of the $\mathbb{Z}$-index map
is an approach starting from the 2D limit, i.e., 
an approach following the evolution of edge properties
in the process of stacking more and more layers.
The conspicuous even-odd feature emergent
in the middle of the mosaic-like pattern formed at the merger of
the ${\cal N}>0$ and ${\cal N}<0$ sides
was understood from this point of view.
While, in the previous subsection
the mosaic pattern on the $\mathbb{Z}_2$-side has been discussed
from the viewpoint of surface/bulk picture,
i.e., in terms of the damped-oscillatory pattern of the top-bottoms surface wave function,
penetrating through the bulk of the film to attain the opposite surface,
leading to peculiar oscillatory (mosaic) pattern of the hybridization gap.
Since the mosaic pattern on the $\mathbb{Z}_2$-side reduces to that of the $\mathbb{Z}$-side 
in the limit of $t_z \rightarrow 0$, it is natural to assume
that the mosaic pattern on the $\mathbb{Z}$-side
can also be understood from such a surface/bulk point of view.
However, one may realize in a moment that this might not be the case,
since on the $\mathbb{Z}$-side, i.e., in the CI/WSM film
the top-bottoms surfaces are always gapped,
so that {\it a priori} there is not even a starting point for developing a similar
surface/bulk picture. 

This apparent contradiction can be resolved as follows.
The penetration of the top-bottom surface wave function in the WTI/STI model 
becomes deeper as the magnitude of SOC hopping $t_z$ is diminished.
{\it In the limit of $t_z \rightarrow 0$,
the surface wave function evolves continuously into a bulk wave function
associated with a pair of new bulk Dirac points.}

Let us start with the STI $(1;000)$ phase 
represented by the point:
$(m_0/b_\parallel, b_z/b_\parallel)=(2,1)$ in Fig.~\ref{PD}(a).
On top and bottom surfaces of the relatively thick film (or slab)
placed normal to the $z$-axis,
the surface Dirac cone appears at the $\bar{\Gamma}$-point:
$(k_x, k_y)=(0,0)$.
The phase boundaries of Fig.~\ref{PD}(a)
correspond to either of the gap closing points of Eq.~(\ref{H_TI})
at one of the eight time reversal invariant momenta (TRIM)
in the 3D BZ: $k_\mu = 0, \pi$.
Conversely, as far as $t_\mu$ is finite, 
gap closing of Eq.~(\ref{H_TI}) occurs only at such 8 TRIM.
However, in the limit of $t_z \rightarrow 0$,
chance for a new type of gap closing arises, 
i.e.,
at $k_z\neq 0, \pi$;
indeed, at
\begin{equation}
k_z=\pm \arccos\left({m_0\over b_z} - 2{b_\parallel\over b_z}\right) \equiv \pm k_0,
\end{equation}
or at the value of $k_z$ satisfying
\begin{equation}
{m_0\over b_\parallel}-2 = {b_z\over b_\parallel}\cos k_z.
\label{k0}
\end{equation}
In the isotropic case: $b_z/b_\parallel =1$ 
the STI $(1;000)$ phase corresponds to the range: $1<m_0/b_\parallel<3$;
therefore, the left-hand side of the equation is between 1 and $-1$.
Thus, $k_0$ satisfying the condition (\ref{k0}) always exists in the interval: $[-\pi, \pi]$.
In the limit of $t_z \rightarrow 0$,
the top and bottom surface wave functions reduce to
the following bulk wave functions associated with these new Weyl points 
that appear at $k_z = \pm k_0$, i.e.,
\begin{equation}
\psi(z) = A \sin (k_0\pm q)z,
\end{equation}
where $q$ is a dips 
from the Weyl point determined by the boundary condition: 
$\psi(N_z+1)=0$.
\cite{mayu1}

The number ${\cal N}_D$ of surface Dirac cones that appear on top and bottom
surfaces of the slab
differ in different phases of Fig.~\ref{PD}(a).
The above argument shows that the number ${\cal N}_W$
of Weyl pairs that appear in 
each of the WSM phase in Fig.~\ref{PD}(b)
coincide with the number ${\cal N}_D$ of the corresponding WTI/STI phase:
\begin{equation}
{\cal N}_D={\cal N}_W.
\end{equation}

\begin{figure*}[tbp]
\begin{tabular}{l}
\includegraphics[width=120mm, bb =0 0 761 450]{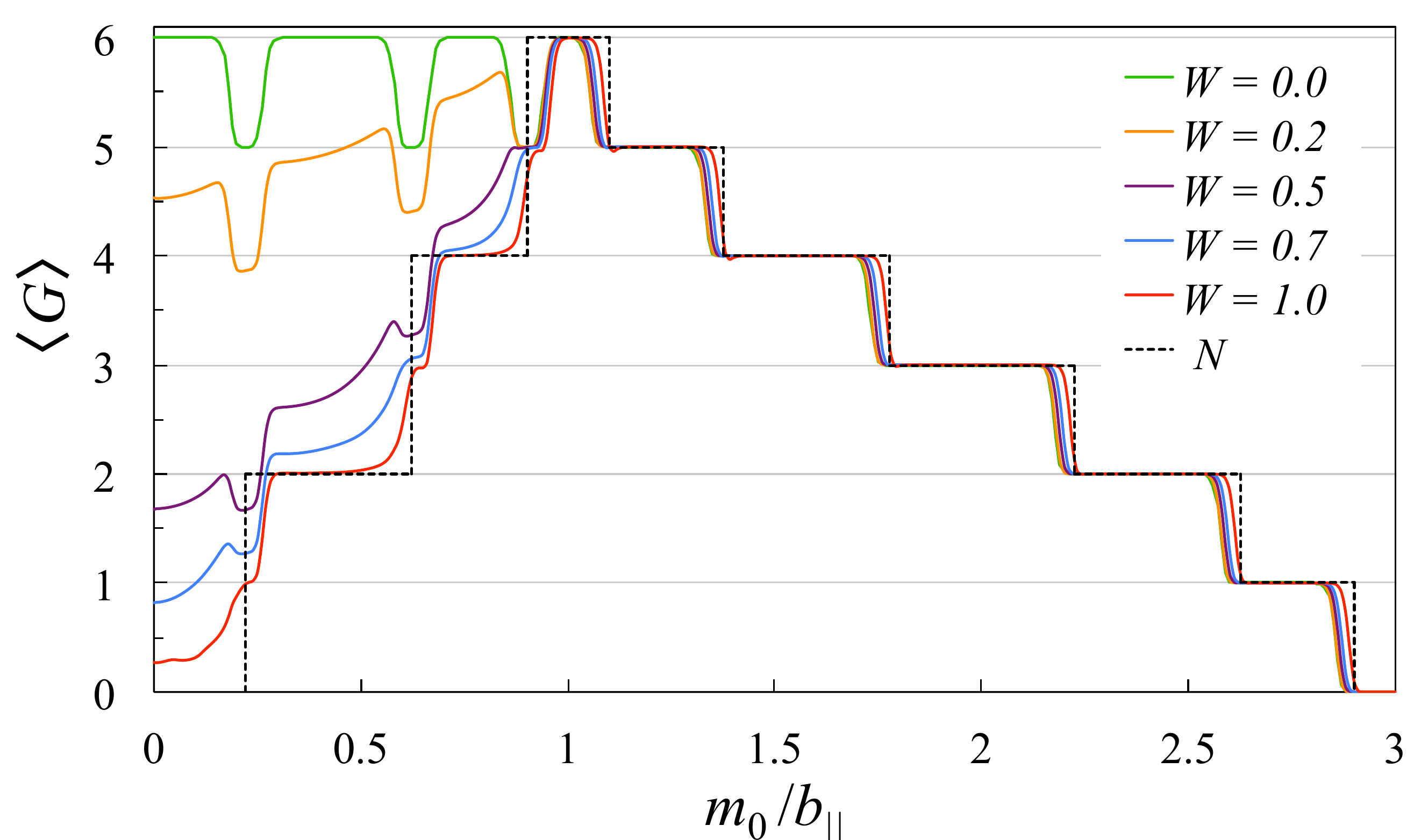}
\end{tabular}
\vspace{-2mm}
\caption{
 Conductance for CI/WSM thin-films.
 The two-terminal conductance $G$ for a system of
$N_x=200$, $N_y=50$, $N_z=6$
is plotted
in units of $e^2/h$ at different values of disorder strength $W$.
The dotted line represents 
the change of bulk topological $\mathbb{Z}$-index,
extracted from Fig.~\ref{zmap}(b).
}
\label{cmap_Lz6}
\end{figure*}

\section{Effects of disorder and relation to experiments}

The above $\mathbb{Z}$- and $\mathbb{Z}_2$-index maps are closely related to 
conductance of the system.
\cite{cmap}
 By calculating the conductance,
we can also discuss 
how robust
the characteristic features such as the stripe vs.~mosaic patterns
emergent in the $\mathbb{Z}$- and $\mathbb{Z}_2$-index maps
are in the presence of disorder.
 Here we 
compare closely
the topological 
(i.e., $\mathbb{Z}$- or $\mathbb{Z}_2$-) index
vs.~conductance 
in the CI/WSM case.
A similar study of conductance maps has been done
in Ref. \onlinecite{cmap}
in the case of WTI/STI thin-films.
The advantage of the conductance map is that one can equally consider the case of
zero and finite disorder on the same footing. 

Here, we focus on the behavior of the two-terminal conductance $G$
calculated in the transfer matrix approach
employed in Ref. \onlinecite{cmap}.
$G$ has been calculated for a rectangular
: $N_x\times N_y\times N_z$
with the three axes $x$, $y$, $z$, placed respectively in the direction of 
$x$: conduction, $y$: width and $z$: stacking.
Here, we set $b_z/b_\parallel = 1$.
The conductance has been studied in detail for a system of $N_x=200$, $N_y=50$
and several different $N_z$.
In Fig.~\ref{cmap_Lz6}
the calculated conductance $G$ at $N_z=6$ 
at the varying ratio of mass and Wilson parameters ($m_0/b_\parallel$)
is shown.
Different curves correspond to different strength of disorder $W$;
here we consider site disorder,
i.e., random potential on each site, 
obeying the uniform distribution $[-W/2, W/2]$.
For curves with $W\neq 0$
an average over independent configurations of disorder is taken for 10 samples.
In Fig.~\ref{cmap_Lz6}
the calculated conductance curves are plotted against a dotted line;
the latter represents
the change of the bulk topological $\mathbb{Z}$-index,
extracted from Fig.~\ref{zmap}(b).

For sufficiently large $N_x$ and $N_y$,
the $\mathbb{Z}$-index map coincides identically with the {\it Hall} conductance ($G_H$) map.
 On the other hand,
here we consider the two-terminal conductance $G$.
 In Fig.~\ref{cmap_Lz6} let us first note that
the step structure on the outer side of the $\mathbb{Z}$-index map
starting at $m_0/b_\parallel = 3.0$
is well reproduced in the conductance calculation.
This is so, since in the regime of $m_0/b_\parallel\gtrsim 1$,
only edge modes of the same chirality are 
responsible for conduction.
We then note that in the central mosaic region ($m_0/b_\parallel\lesssim 1$),
the discrepancy of the two maps arises.
In this region
the Chern number $\cal N$ (or equivalently $G_H$) decreases
as $m_0/b_\parallel$ decreases,
forming the inner staircase of the map,
while in conductance map,
$G$ is kept at the maximal saturated value
in the clean limit $W=0$.
This is because $G$ measures only the number of transmitting
channels; here they are edge modes circulating around the periphery of the film-bar shaped 
sample. 
{\it By its nature}, $G$ is 
insensitive to the chirality of edge modes.
\cite{KaneMele_QSH}
If ${\cal N}_+$ (${\cal N}_-$) edge modes of the $+$ ($-$) chirality
are available for conduction,
the Hall conductance is given as
\begin{equation}
G_H = ({\cal N}_+ - {\cal N}_-) {e^2\over h} = {\cal N}{e^2\over h},
\end{equation}
while the two-terminal conductance becomes
\begin{equation}
G = ({\cal N}_+ + {\cal N}_-) {e^2\over h}.
\end{equation}
 In the clean limit $W=0$,
a few (here, in the case of $N_z=6$, three) pairs of counter-propagating modes are
circulating, practically without backscattering (since the bulk is gapped).
 Only at the value of $m_0/b_\parallel$ corresponding to the steps,
the bulk becomes gapless, 
allowing for weak backscattering through the bulk,
which explains small deviation of $G$ at $W=0$
from the maximal value $G=6$
in the vicinity of these values of $m_0/b_\parallel$.

As disorder is increased,
the value of $G$ decrease to the value
predicted by the $\mathbb{Z}$-index map
in the central mosaic area, corresponding to the WSM phase.
This is because of the backscattering between the counter propagating chiral states due to disorder, 
resulting in $G = |{\cal N}_+ - {\cal N}_-| {e^2\over h}$.
The typical mosaic-stripe pattern characteristic to WSM thin-films is a feature,
albeit predicted by the $\mathbb{Z}$-index map,
manifesting
only on addition of finite disorder,
and in this sense,
may be called an emergent property 
induced by the interplay of nontrivial topological nature of WSM and disorder.

Let us note here that
a step-like change of the conductance 
somewhat reminiscent of the ones shown in Fig.~\ref{cmap_Lz6}
has been reported in a slightly
different context.\cite{burkov-1,HgCrSe} 
However,
the step structure reported in these references
appears as varying the thickness of the film, and
is associated with quantization of $k_z$ due to the finite thickness.
On the other hand, the step structure shown in Fig.~\ref{cmap_Lz6} 
is due to gradual destruction of the CI phase
in the process of becoming a WSM, and occurs at a fixed $N_z$.
In this sense
the nature of the step structure shown in Fig.~\ref{cmap_Lz6}
is qualitatively different from the ones reported in Refs.~\onlinecite{burkov-1} and \onlinecite{HgCrSe}.

\section{Concluding remarks}

In this paper,
a comparative study of the 
CI/WSM and WTI/STI type topological insulators and semimetals
has been done 
from the viewpoint of thin-film construction.
We introduced a correspondence relation between
the two classes of models,
based on a specific model defined in Sec.~II
and the corresponding phase diagram shown in Fig.~\ref{PD}.
In particular, we have argued that the counter part of
STI and WTI in the 
time-reversal symmetry (TRS) broken class is, respectively, 
WSM and CI phase. 
We have also demonstrated that
STI can be regarded as partially broken WTI,
and in the same way
WSM can be regarded as partially broken CI.
Much of the subsequent analysis was developed, based on this parallelism.

Here, let us comment on how generic this correspondence relation could be.
Naturally,
the exact correspondence of the two phase diagrams shown in Fig.~\ref{PD}
is due to a specific choice of our model.
Yet, the comparison and the analogy
developed in this analysis could be more generic and valid for a broad class of
TRS broken WSM phases. 
Starting with the time-reversal symmetric limit, 
let us imagine adding a perturbation that breaks TRS. 
Then, the concept of STI and WTI is invalidated since TRS is no longer existent. 
Still, in some cases 
a specific type of WTI phase is not much affected by such a perturbation, 
but simply replaced with a CI phase.
On the other hand, 
surrounding STI phases disappear immediately as the TRS breaking perturbation turned on. 
\cite{burkov-1}
They are considered to be generically transformed to a WSM phase.
Then, it is likely that one ends up, at least, qualitatively, 
with the same situation as described by the model employed here.
The situation would be much different if one considers
WSM of the inversion symmetry (IS) breaking type.
\cite{herring}
In the limit of both TRS and IS preserved,
STI and WTI phases are separated by a Dirac semimetal (DSM) line.
Adding IS breaking perturbation does not generically destroy
STI and WTI phases,
while a finite WSM phase replacing the DSM line appears in between. 
Such a finite WSM region appears also between STI and OI phases.

To reveal the dimensional crossover of topological signatures in the two class of models,
we have focused on the thin-film construction of the two types of models.
To quantify the crossover,
we have focused on the dimensional crossover of topological signatures in the two class of models.
To quantify this,
we have introduced, and then evaluated the 2D topological indices
of $\mathbb{Z}$- and $\mathbb{Z}_2$ types,
characterizing the system.
We mapped such topological indices ($\mathbb{Z}$- and $\mathbb{Z}_2$-index maps)
as a function of the gap parameter and film thickness,
to reveal characteristic global patterns, such as stripe or mosaic patterns,
encoding the topological properties of the system in the 3D limit.
The nature of the two characteristic patterns are discussed and
understood from two complementary points of view:
surface/bulk vs.~edge pictures.
The similarity and relation of the $\mathbb{Z}$- and $\mathbb{Z}_2$-type
index maps are discussed and clarified.

It was shown that
in the limit of $t_z \rightarrow 0$
essentially all the phase boundaries of the $\mathbb{Z}_2$-index map
coincide with those of the $\mathbb{Z}$-index map.
We have demonstrated
how
surface helical Dirac cones emergent and protected on the surface of an STI
evolve and eventually transform into
a pair of Weyl cones in the bulk of the WSM.
The number of surface Dirac cones ${\cal N}_D$ 
on the surface of an STI,
or of WTI in some cases,
is identical to the number of Weyl cone pairs ${\cal N}_W$
lurking in the bulk of the corresponding WSM.

Finally,
to check the robustness of our findings against disorder
and their relevance to experiments
we have also made a close comparison of
the topological $\mathbb{Z}$-index map (in the clean limit)
with the corresponding conductance map
at varying strength of disorder. 
Compared with the previously studied case of WTI/STI thin-films,
\cite{cmap}
in which the correspondence
between the two-terminal conductance map at finite disorder
with the topological $\mathbb{Z}_2$-index map
was apparent,
here in the present CI/WSM case studied in detail
in the previous section
the correspondence between the conductance and $\mathbb{Z}$-index maps
is only partial in the clean limit,
since as discussed in the previous section,
the $\mathbb{Z}$-index map should show a perfect correspondence 
with the Hall conductance map.
Yet, we have argued that
the two-terminal conductance at varying strength of disorder
those features that we have revealed in the study of $\mathbb{Z}$-index map
appears as an emergent property,
which manifests 
only at some finite strength of disorder.

\begin{acknowledgments}
The authors acknowledge Yositake Takane for useful comments on the manuscript.
This work was supported by JSPS KAKENHI Grant Numbers 15H03700, 15K05131, 24000013, 
and 
16J01981.
YY is supported by Grant-in-Aid for JSPS Fellows and by Grant No. 15J06436.
\end{acknowledgments}

\appendix

\begin{figure*}[tbp]
\begin{tabular}{c}
(a)
\includegraphics[width=150mm, bb =0 0 700 202]{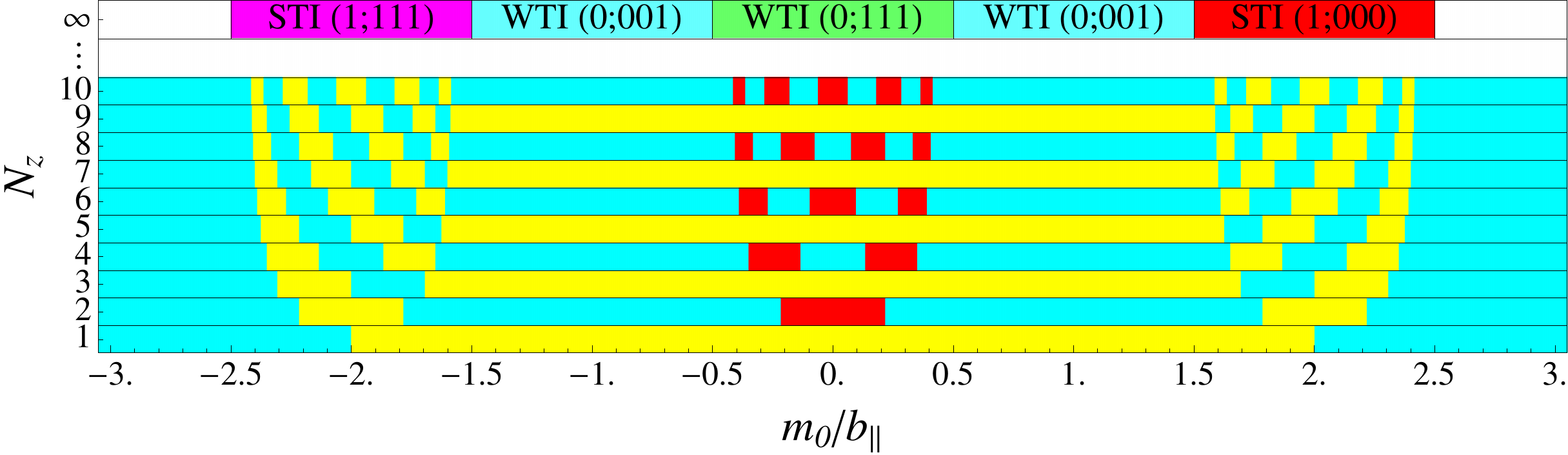}
\\
(b)
\includegraphics[width=150mm, bb =0 0 700 202]{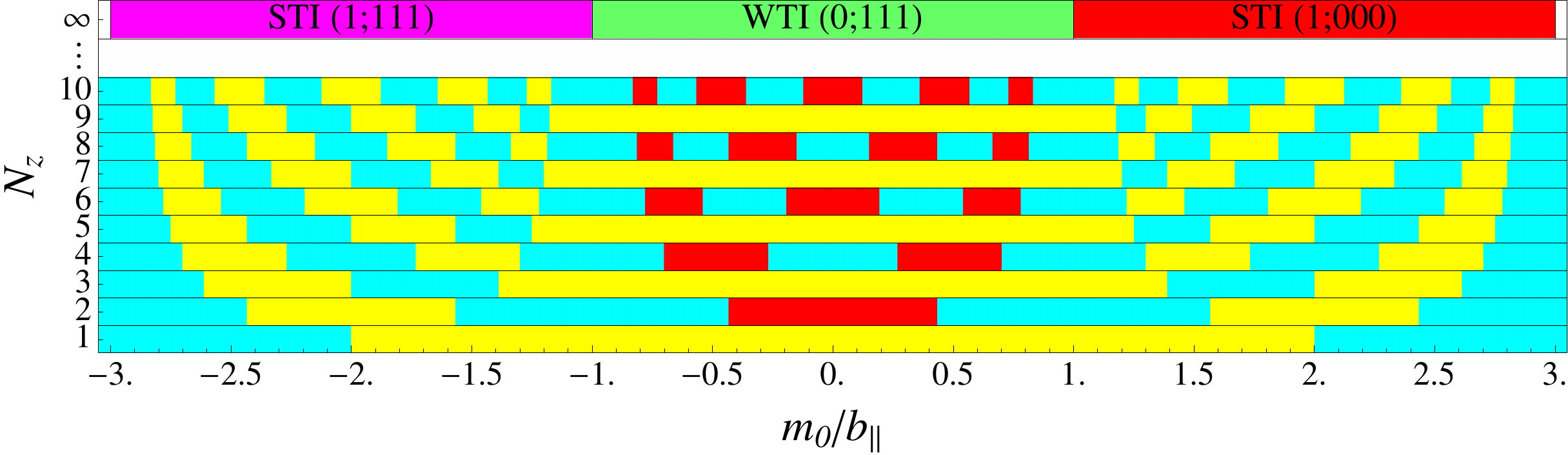}
\end{tabular}
\vspace{-2mm}
\caption{
$\mathbb{Z}_2$-index map with 2D weak-like indices \cite{cmap}.
The yellow (cyan/red) region corresponds to the range of $\nu=1$ ($\nu=0$),
corresponding to red (white) regions in Fig.~\ref{z2map}.
The red region represents the 2D weak phase.
}
\label{z2map_w}
\end{figure*}

\section{Surface/bulk vs. edge picture for the $\mathbb{Z}_2$-index map}

\subsection{Nature of the mosaic pattern: the surface/bulk mechanism}



On the isotropic line: $b_z=b_\parallel$,
the range of parameter: $1 < m_0 / b_\parallel < 3$ corresponds to an STI phase 
with a surface Dirac cone at the $\bar{\Gamma}$-point.
\cite{mayu1}
Let us examine in this case whether
hybridization of the top and bottom surface wave functions
in the thin-film geometry
leads to inversion of the valence and conduction bands,
characteristic to an effective QSH state.

The surface wave function in the semi-infinite geometry
takes the following form:
\begin{equation}
\label{wf}
\psi (z) \propto \rho_1^z - \rho_2^z,
\end{equation}
where
$\rho_{1,2}$ are two solutions to a characteristic equation
associated with the surface wave function at $E=0$,
i.e.,
\begin{equation}
\label{rho12}
\rho_{1,2} = {\pm\sqrt{D}-(m_0 - 2 b_\parallel ) \over b_z - t_z},
\end{equation}
with the discriminant $D$ given as 
\begin{equation}
D=(m_0 - 2 b_\parallel )^2+t_z^2-b_z^2.
\end{equation}
The double sign in the numerator
defines, respectively,
$\rho_1$ and $\rho_2$.
Here,
the system is assumed to be extended, e.g., on the $z>0$ side
with a surface at $z=0$.
To be consistent with this boundary condition
the surface wave function (\ref{wf}) is constructed in such a way that
it vanishes at $z=0$: $\psi (0) = 0$,
but decays exponentially as $z\rightarrow\infty$.
Such a surface solution is possible,
when the magnitude of two solutions $\rho_{1,2}$ in Eq.~(\ref{rho12})
are both smaller than unity: $|\rho_{1,2}|<1$,
which is indeed possible in the present case:
$1 < m_0 / b_\parallel < 3$.

It can be shown that
the sign and magnitude of the energy gap $\Delta E$ of a TI thin-film 
of a thickness $N_z$
is related to those of the surface wave function $\psi (z)$
of an auxiliary semi-infinite system at the depth of the film thickness $N_z$,
or more directly to $\rho_{1,2}$ as
\cite{mayu2}
\begin{eqnarray}
\label{gap}
\Delta E (N_z)
&=& E_c - E_v 
\nonumber \\
&=& A\left( \rho_1^{N_z+1} - \rho_2^{N_z+1}\right)
\nonumber \\
&\propto& \psi (N_z+1), 
\end{eqnarray}
where $A$ is a positive constant.
When $\Delta E <0$ ($\Delta E >0$)
the TI thin-film realizes an effective 2D QSH
(ordinary) insulator.
\cite{shen_NJP}

It will be convenient
to consider the cases of (i) $t_z < b_z$ and (ii) $t_z > b_z$ 
separately, since the profile of the $\mathbb{Z}_2$-index map changes qualitatively,
depending on the sign of $t_z - b_z$.
If (i) $t_z < b_z$,
the two solutions $\rho_{1,2}$ may have an imaginary part since $D$ can be negative;
in this case $\rho_{1,2}$
becomes a pair of conjugate complex numbers [see Eq.~(\ref{rho12})].
Then, the corresponding surface wave function $\psi (z)$ 
becomes oscillatory,
showing a damped oscillation, 
i.e., the wave function changes its sign 
in a somewhat irregular way
as it penetrates into the bulk.
When the thickness $N_z$ of the film is chosen such that
$\psi (N_z+1)<0$,
the hybridization gap $\Delta E$ becomes negative [see Eq.~(\ref{gap})],
implying that the film is in the QSH phase.
The corresponding $\mathbb{Z}_2$-index map shows a mosaic pattern.

If (ii) $t_z > b_z$, on the other hand,
the two solutions $\rho_{1,2}$ are always real since $D$ is always positive.
They are also positive and negative value solutions:
$\rho_1 = \rho_+ >0$ and $\rho_2 = - \rho_- <0$.
In this case
the corresponding surface wave function is overdamped.
Yet, the wave function can still be oscillatory in some cases.
Let us recall that
if (a) $m_0 - 2 b_\parallel > 0$,
the system is on the OI side in the 2D limit.
Eq.~(\ref{rho12}) implies that
$\rho_+ > \rho_-$ in this case,
therefore,
the corresponding surface wave function in the form of Eq.~(\ref{wf})
is always positive.
The resulting $\mathbb{Z}_2$-index map shows a trivial pattern.

If (b) $m_0 - 2 b_\parallel < 0$, on the other hand,
the system is on the QSH side in the 2D limit.
In this case the weight of the two solutions are inverted,
i.e., the opposite inequality, $\rho_+ < \rho_-$ stands.
Then,
the corresponding surface wave function 
\begin{equation}
\label{wf_2}
\psi (z) = \rho_+^z - (-\rho_-)^z
\end{equation}
shows an alternating sign. 
In the $\mathbb{Z}_2$-index map, an alternating series of QSH and OI layers appear,
forming a peculiar stripe pattern.

\subsection{Crossover from mosaic to stripe pattern: complementarity of the two pictures}

When $t_z < b_z$, there exists an interval:
\begin{align}
\label{D<0}
 \! 2 
 -\sqrt{
    \left({b_z \over b_\parallel}\right)^{\!\!2} 
    \!-\! \left({t_z \over b_\parallel}\right)^{\!\!2}
  } 
 < {m_0\over b_\parallel} 
 <  2
 +\sqrt{
    \left({b_z \over b_\parallel}\right)^{\!\!2} 
    \!-\! \left({t_z \over b_\parallel}\right)^{\!\!2}
  },
\end{align}
in which
damped oscillation of the surface wave function
leads to a mosaic pattern of the $\mathbb{Z}_2$-index map.
Note that the interval Eq.~(\ref{D<0}) 
fits inside the window of the STI phase: $1<m_0/b_\parallel<3$.
The interval Eq.~(\ref{D<0}) corresponds to the range of parameters in which 
$D<0$, so that
$\rho_{1,2}$ become a pair of conjugate complex numbers,
resulting in a damped oscillation of the surface wave function.
The corresponding $\mathbb{Z}_2$-index map shows a mosaic pattern.
Beyond the interval Eq.~(\ref{D<0}) 
and inside the STI window
the surface wave function shows a feature
analogous to the case of $t_z > b_z$, 
which means that a stripe pattern appears 
on the side of $m_0 - 2 b_\parallel < 0$.
To be precise,
a mosaic pattern appears only when $N_z$ is sufficiently large
even in the interval (\ref{D<0}).
Recall that the film thickness $N_z$ is discretized to be
an integer multiple of the layer separation $a$,
which is here chosen to be unity.
Due to this discreteness of $N_z$,
evolution of the $\mathbb{Z}_2$-index at small $N_z$
is superficially periodic in spite of the damping.
The resulting pattern is also
practically indistinguishable from a perfect stripe pattern.
Aperiodicity becomes manifest,
however, at finite $N_z$ and a crossover from a mosaic to a stripe pattern occurs.
The boundary between the mosaic and stripe pattern
approaches an asymptotic line in the limit of large $N_z$,
corresponding to either of the extremities of the interval (\ref{D<0}).

The interval (\ref{D<0}) appears as a finite range of $m_0/b_\parallel$
in the STI window: $1<m_0/b_\parallel<3$
in the range: $0<t_z < b_z$.
In the limit of $t_z \rightarrow 0$
the interval dominates the entire STI window;
a mosaic pattern becomes predominant in the STI region.
The interval (\ref{D<0}) vanishes, on the other hand,
as $t_z$ approaches $b_z$;
i.e., the mosaic pattern disappears at $t_z = b_z$.
The situation is unchanged
even when $t_z$ is further increased.

Such an observation shows that 
the extremity of the STI window
toward the boundary to the neighboring WTI phase: 
$-1<m_0/b_\parallel<1$,
is necessarily be ended by a stripe pattern.
The origin of this stripe pattern is 
a particular oscillation pattern of the surface wave function in the form of Eq.~(\ref{wf_2}).
Last but not least,
this stripe pattern continues on the WTI side
beyond the boundary at $m_0/b_\parallel = 1$.
This is because
the surface states at $\bar{\Gamma}$ on two surfaces of the film
is no longer existent;
there are still two Dirac cones on the WTI side
but they appear at $\bar{X}=(\pi,0)$ and $\bar{Y}=(0,\pi)$,
and no more at $\bar{\Gamma}$.
Since there is no more surface state at $\bar{\Gamma}$,
no more gap inversion occurs at $\bar{\Gamma}$.
There still can occur band inversions at $\bar{X}$ and $\bar{Y}$, 
but at least on the isotropic line
they occur simultaneously at the two points;
therefore, irrelevant to the $\mathbb{Z}_2$-index map shown in Fig.~\ref{z2map}.
In Fig.~\ref{z2map_w}
2D weak-like indices
\cite{YIFH}
are added to the 
the $\mathbb{Z}_2$-index map shown in Fig.~\ref{z2map}.
This introduces an additional mosaic-like structure
in the WTI $(0;111)$ region
stemming from the band inversion at $\bar{X}$ and $\bar{Y}$ points.

We have so far seen that
the stripe pattern in the WTI $(0;111)$ region 
as shown in Fig.~\ref{z2map}
can be regarded as
a {\it heritage} of the same pattern at an extremity of the neighboring
STI phase.
The stripe pattern in the WTI phase is also sometimes 
understood from the viewpoint of edge wave functions 
emergent on narrow side surfaces of the film.
\cite{mayu1}
In this second point of view only edge/side-surface
properties are involved,
while the present scenario involves only
the penetration of the top-bottom-surface wave function into the bulk.
This complementarity of the two pictures is the hallmark of
all topological quantum phenomena,
often referred to as the {\it bulk-boundary} correspondence.
\cite{HG} 

In the above surface/bulk point of view,
the stripe pattern in the WTI $(0;111)$ region 
as shown in Fig.~\ref{z2map}
can be regarded as
a {\it heritage} of the same pattern at an extremity of the neighboring
STI phase.
In Sec. III A
we have seen
that the stripe pattern in the WTI phase can be also sometimes 
understood from the viewpoint of edge wave functions 
emergent on narrow side surfaces of the film.
This complementarity of the edge vs. surface/bulk 
pictures can be regarded as a manifestation of the
bulk-boundary correspondence,
\cite{HG,Wen} 
which is believed to be the hallmark of all topological properties.

\bibliography{PCI2-20}

\end{document}